\documentclass[12pt]{article}
\usepackage[utf8]{inputenc}
\usepackage[a4paper, top=2.5cm, bottom=2.5cm, left=2.2cm, right=2.2cm]
{geometry}
\usepackage{amsmath,amssymb,amsfonts,bm,amscd,mathtools}
\usepackage[bbgreekl]{mathbbol}
\usepackage{graphicx}
\usepackage{caption, subcaption}
\usepackage{comment}
\usepackage{cite}
\usepackage{float}
\usepackage[titletoc,title]{appendix}
\usepackage{physics}
\usepackage{tikz}
\usepackage{tikz-cd}
\usetikzlibrary{cd}
\usepackage{array}
\usepackage[hidelinks]{hyperref}

\usepackage{tocbasic}
\DeclareTOCStyleEntry[
  beforeskip=.4em plus 1pt,
  pagenumberformat=\textbf
]{tocline}{section}

\newtheorem{definition}{Definition}
\newtheorem{conjecture}{Conjecture}[section]

\newcommand{\Z}{{\mathbb Z}}

\newcommand{\C}{{\mathbb C}}
\newcommand{\Q}{{\mathbb Q}}

\def\tilde{\widetilde}

\renewcommand{\hat}{\widehat}

\def\be{\begin{equation}}
\def\ee{\end{equation}}

\begin{document}

\title{\huge \bf $c_{\text{eff}}$ for 3d $\mathcal{N}=2$ theories}

\author{Sergei Gukov$^{a,b}$, Mrunmay Jagadale$^c$}





\date{%
  $^a$Richard N. Merkin Center for Pure and Applied Mathematics, California Institute of Technology, Pasadena, CA 91125, USA\\%
  $^b$Dublin Institute for Advanced Studies, 10 Burlington Rd, Dublin, Ireland\\
  $^c$Walter Burke Institute for Theoretical Physics,  California Institute of Technology, Pasadena, CA 91125, USA\\[2ex]
}

\maketitle

\begin{abstract}
Based on the observed behavior of the superconformal index in three-dimensional $\mathcal{N}=2$ theories, we propose a quantity that can be considered as an analogue of the ``effective central charge.'' We discuss the general properties of this quantity and ways of computing it in a variety of different theories, including simple Lagrangian theories as well as more interesting strongly coupled examples that come from 3d-3d correspondence.
\end{abstract}
\newpage
\tableofcontents




\section{Introduction}

Effective measures of degrees of freedom in Quantum Field Theories can be especially useful in understanding RG flows that go way outside the perturbative regime. Each time there is a candidate for such a quantity --- often called a ``$c$-function'' --- one faces a variety of questions: Is it defined away from fixed points of the RG flow? If so, is the flow a gradient flow?

The answer to these questions very much depends on $d$, the space-time dimension. The state of affairs is perhaps most satisfying for $d=2$ since, in this case, RG-flows are indeed gradient flows with respect to the $c$-function~\cite{Zamolodchikov:1986gt}. The next best case is in $d=4$ where a weaker version of the $c$-theorem holds, in the sense $c_{UV} > c_{IR}$, but it is not known whether 4d RG flows are gradient flows \cite{Cardy:1988cwa,Barnes:2004jj,Komargodski:2011vj}. (In four dimensions, the relevant function is often called $a$; we still refer to it as a $c$-function here for uniformity of notations.)

In both $d=2$ and $d=4$, these measures of degrees of freedom can be defined as conformal trace anomaly coefficients.
Unfortunately, this type of definition is only limited to even values of $d$ and, indeed, in $d=3$, the search for a suitable candidate of a $c$-function --- that is monotonic along RG flow and is stationary at the fixed points --- has been considerably more challenging, see {\it e.g.} \cite{Jafferis:2011zi,Klebanov:2011td,Casini:2012ei,Klebanov:2012va}. That's why in this paper, we wish to focus on three-dimensional QFT's, or QFT$_3$ for short.

The reason $c$-functions measure the effective number of degrees of freedom is that, at least in $d=2$ and $d=4$ where they are better understood, they can be formulated directly in terms of the spectrum of the theory, namely expressing the growth of states. For example, in $d=2$, if $a_n$ is the number of states with energy $n$, then
\be
a_n \; \sim \; \exp 2\pi \sqrt{\frac{1}{6} c_{\text{eff}} \, n}
\qquad
\text{as}
\quad
n \to \infty
\label{Cardyan}
\ee
for some constant $c_{\text{eff}}$ \cite{Cardy:1986ie}. Equivalently, it can be formulated as a statement about the behavior of the generating function
\be
\chi (q) \; = \; \sum_n a_n q^{n + \Delta}
\ee
where $\Delta$ is the ground state energy. The starting point of this paper is a curious observation that \eqref{Cardyan} holds in theories one dimension higher, namely in SCFT$_3$.

\begin{conjecture}\label{conj:basic}
In every three-dimensional superconformal field theory with $\mathcal{N}=2$ supersymmetry, the spectrum of supersymmetric (BPS) states obeys \eqref{Cardyan}. In other words, the superconformal index or, equivalently, $S^{2}\times_{q}S^{1}$ partition function,
\be
\mathcal{I}(q) := \Tr_{\mathcal{H}_{S^{2}}}\left[(-1)^{F}q^{R/2+J_{3}}\right]= Z(S^{2}\times_{q}S^{1}) = \sum_{n} a_{n} q^{n}
\label{SSindex}
\ee
enjoys \eqref{Cardyan}.
\end{conjecture}

Although this behavior has been observed in 3d-3d correspondence (see {\it e.g.} \cite{Cheng20193dM,Ekholm:2021irc,Cheng:2022rqr}), to the best of our knowledge, it has never been proposed as a general property of 3d SCFTs with $\mathcal{N}=2$ supersymmetry. What makes this observation interesting is that the number of non-BPS states in a general CFT$_3$ grows much faster. Namely, in CFT$_d$, we have \cite{Kutasov:2000td}
\be
\log a_n \; \sim \; n^{\frac{d-1}{d}}.
\ee
In particular, for $d=2$ the exponent is $\frac{1}{2}$ and for $d=3$ it is $\frac{2}{3}$. Therefore, while the growth of general (non-BPS) states is controlled by $n^{2/3}$, the supersymmetric (BPS) states grow only as $n^{1/2}$ for large values of $n$. Based on this observation, we can make the following definition.

\begin{definition}\label{ceffdef1}
Assuming Conjecture~\ref{conj:basic}, to any 3d $\mathcal{N}=2$ SCFT we associate a quantity $c_{\text{eff}}$ defined via the asymptotic behavior of superconformal index \eqref{SSindex}:
\be
c_{\text{eff}} \; := \; \frac{3}{2\pi^2} \lim_{n \to \infty} \frac{(\log |a_n|)^2}{n}
\label{ceffdef}
\ee
\end{definition}

Because the superconformal index is invariant along RG flows, we can't conjecture that $c_{\text{eff}}$ defined in this way is decreasing along RG flows. Nevertheless, we hope it can still be a useful measure of the number of degrees of freedom in a 3d $\mathcal{N}=2$ SCFT.

The reason the absolute value in \eqref{ceffdef} is used is that, in the supersymmetric setting, the coefficients $a_n$ may not be all positive. After all, they are counting BPS states with signs.\footnote{This aspect has important consequences, namely the 3d superconformal index is expected to admit a categorification \cite{Gukov:2017kmk}.} In part for this reason, below we will need a slightly refined version of Conjecture~\ref{conj:basic} and Definition \ref{ceffdef1} of $c_{\text{eff}}$.

\begin{conjecture}\label{bpscardy1}
The growth of supersymmetric (BPS) states in a three-dimensional $\mathcal{N}=2$ theory is given by the following asymptotic formula,
\be \label{bpscardy}
a_{n} \sim \mathrm{Re}\left[\exp(\sqrt{\frac{2 \pi^{2}}{3} c_{\text{eff}} \hspace{0.15cm} n }  + 2 \pi i r n )\right],
\ee
where $r\in \Q/\Z $, and $c_{\text{eff}} \in \C$.
\end{conjecture}

\begin{figure}[ht]
    \centering
    \includegraphics[scale=0.5]{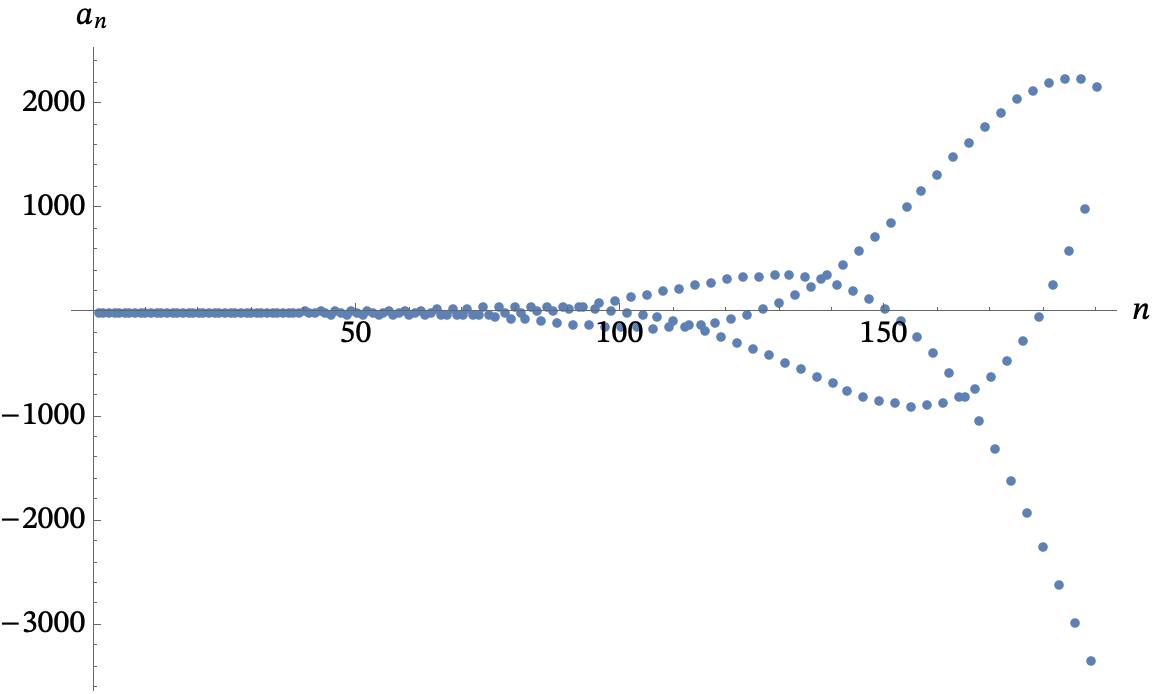}
    \caption{Coefficients of the supersymmetric index in the $\mathcal{N}=2$ Landau-Ginzburg theory with the superpotential $W=\Phi^{3}$.}
    \label{phicubedcoeff}
\end{figure}

This asymptotic formula for the density of BPS states is very reminiscent of the Cardy formula \eqref{Cardyan} for the density of states in two-dimensional conformal field theories. In fact, when $r$ is zero, and $c_{\text{eff}}$ is real and positive, it is exactly the same as the Cardy formula. When $r$ is non-zero, or when $c_{\text{eff}}$ is a generic complex number, the formula \eqref{bpscardy} captures two curious features of the density of BPS states that we call ``branching'' and ``oscillations.'' When $r= \frac{\ell}{k} \neq 0$ with $\ell$ and $k$ relatively prime integers, the density of BPS states branches into $k$ branches. On the other hand, when the imaginary part of $\sqrt{c_{\text{eff}}}$ is non-zero, the density of BPS states oscillates. These two curious features can be seen already in one of the simplest examples of three-dimensional $\mathcal{N}=2$ SCFTs, namely in the theory of a chiral superfield $\Phi$ with the cubic superpotential $W=\Phi^{3}$. As illustrated in Figure~\ref{phicubedcoeff}, the coefficients $a_n$ in this theory exhibit oscillations and form three distinct branches.

In the original derivation of \eqref{Cardyan} by Cardy \cite{Cardy:1986ie}, the $SL(2,\mathbb{Z})$ modular symmetry group of the underlying torus $T^2$ played a key role. If one had a similar symmetry in three dimensions, the derivation of \eqref{bpscardy} would be much easier, and we would not need to state it as a conjecture. As mentioned above, 3d superconfomal index \eqref{SSindex} can be understood as a supersymmetric partition function on $S^{2}\times_{q}S^{1}$. And, although it produces a $q$-series reminiscent of 2d CFT characters or elliptic genera, it lacks any obvious modular properties because $SL(2,\mathbb{Z})$ is not a symmetry of $S^{2}\times_{q}S^{1}$. Nevertheless, the modular group does play a more subtle role in 3d $\mathcal{N}=2$ theories,\footnote{It controls the spectrum of supersymmetric line operators and twisted partition functions \cite{Gukov:2016gkn} in a way that all such relations are mutually compatible, and also compatible with the modular properties of BPS partition functions \cite{Cheng20193dM}.} which can be traced to embedding of $T^2$ into $S^2 \times S^1$ as a central surface in a genus-1 Heegaard splitting. A similar property is shared by $D^{2}\times_{q}S^{1}$ supersymmetric partition function, where $T^2$ is the boundary and its $SL(2,\mathbb{Z})$ is also manifest.

As a result, some supersymmetric partition functions of 3d $\mathcal{N}=2$ theories exhibit a ``universality'' in the sense that their behavior near roots of unity is controlled by semi-classical quantities such as the twisted superpotential \cite{Gadde:2013wq}. Here, we summarize this 3d analogue of modularity in the form of the following conjecture that will be useful to us not only in the analysis of the superconformal index $\mathcal{I}(q)$ but also in its close cousins that encode other types of BPS spectra. In other words, our analysis below will assume the following behavior of the relevant supersymmetric partition functions.

\begin{conjecture}\label{bpscardy2}
Supersymmetric partition functions have the following asymptotic behavior near $q = e^{-2 \pi i r}$ or equivalently near $\tau = - r$, with $r\in \Q/\Z$.
\be \label{singWW}
Z(q) \sim \exp[-\frac{1}{\hbar_{r}} \left( \widetilde{\mathcal{W}}_{r}^{(0)}+\hbar_{r}\widetilde{\mathcal{W}}_{r}^{(1)}+ \hbar_{r}^{2}\widetilde{\mathcal{W}}_{r}^{(2)} +\cdots \right)],
\ee 
where $e^{-\hbar_{r}} = q e^{2 \pi i r} =e^{2\pi i (\tau+r)} $, and $\widetilde{\mathcal{W}}_{r}^{(j)} \in \C$.
\end{conjecture}
Using this conjecture as an intermediate step, we can derive the density of states in Conjecture \ref{bpscardy1} in much the same way as the Cardy formula \eqref{Cardyan} is derived in two-dimensional conformal field theories. In other words, as we explain in more detail in the next section, it reduces the analysis to the study of the asymptotic behavior near the natural boundary $|q|=1$.

We verified Conjectures \ref{bpscardy1} and \ref{bpscardy2} in a large variety of 3d $\mathcal{N}=2$ theories. In Lagrangian theories, such as $\Phi^3$ theory mentioned earlier, the analysis easily follows the strategy outlined above because indices can be explicitly written as sums of basic ingredients made of $q$-Pochhammer symbols, and their asymptotics at roots of unity is well known.

We also considered some 3d $\mathcal{N}=2$ theories whose Lagrangian description is not known at present. A large class of such theories comes from 3d-3d correspondence. As argued in \cite{Chun:2019mal}, one should expect to be able to realize theories $T[M_3]$ as gauge theories with ``non-linear'' matter, {\it i.e.} as Skyrme type models where 3d $\mathcal{N}=2$ chiral multiplets take values in complex group manifolds $G_{\C}$. At present, such a description is not developed. However, in the context of 3d-3d correspondence, there is another strategy available to us that can be very helpful in approaching Conjectures \ref{bpscardy1} and \ref{bpscardy2}. It relies on the fact that, for any 3-manifold $M_3$, theories $T[M_3]$ admit a canonical choice of 2d $\mathcal{N}=(0,2)$ boundary conditions labeled by Spin$^c (M_3)$. This allows us to reduce the analysis of the superconformal index $\mathcal{I}(q)$ to a (sometimes simpler) analysis of a family of BPS partition functions on $D^{2}\times_{q}S^{1}$ with boundary conditions labeled by Spin$^c$-structures. (The reason this strategy does not admit an obvious extension outside 3d-3d correspondence is that general 3d $\mathcal{N}=2$ theories do not have a ``canonical'' set of 2d $\mathcal{N}=(0,2)$ boundary conditions.)

The rest of the paper is organized as follows. In section~\ref{sec:modular}, we imitate Cardy's derivation of \eqref{Cardyan} in the context of 3d $\mathcal{N}=2$ theories, using \eqref{singWW} as an assumption. We then illustrate general ideas and considerations with a concrete example of $\Phi^3$ theory, whose BPS spectrum already appeared in Figure~\ref{phicubedcoeff}. In section~\ref{sec:3d3d}, we turn to examples of 3d $\mathcal{N}=2$ theories, for which Lagrangian description is not known at present. Although we draw such examples from 3d-3d correspondence, one can probably consider other sources. In all instances, we find evidence for Conjectures \ref{bpscardy1} and \ref{bpscardy2} and compute the corresponding values of $c_{\text{eff}}$.


\section{Indices near the unit circle}
\label{sec:modular}

\begin{figure}[ht]
    \centering
    \includegraphics[scale=0.5]{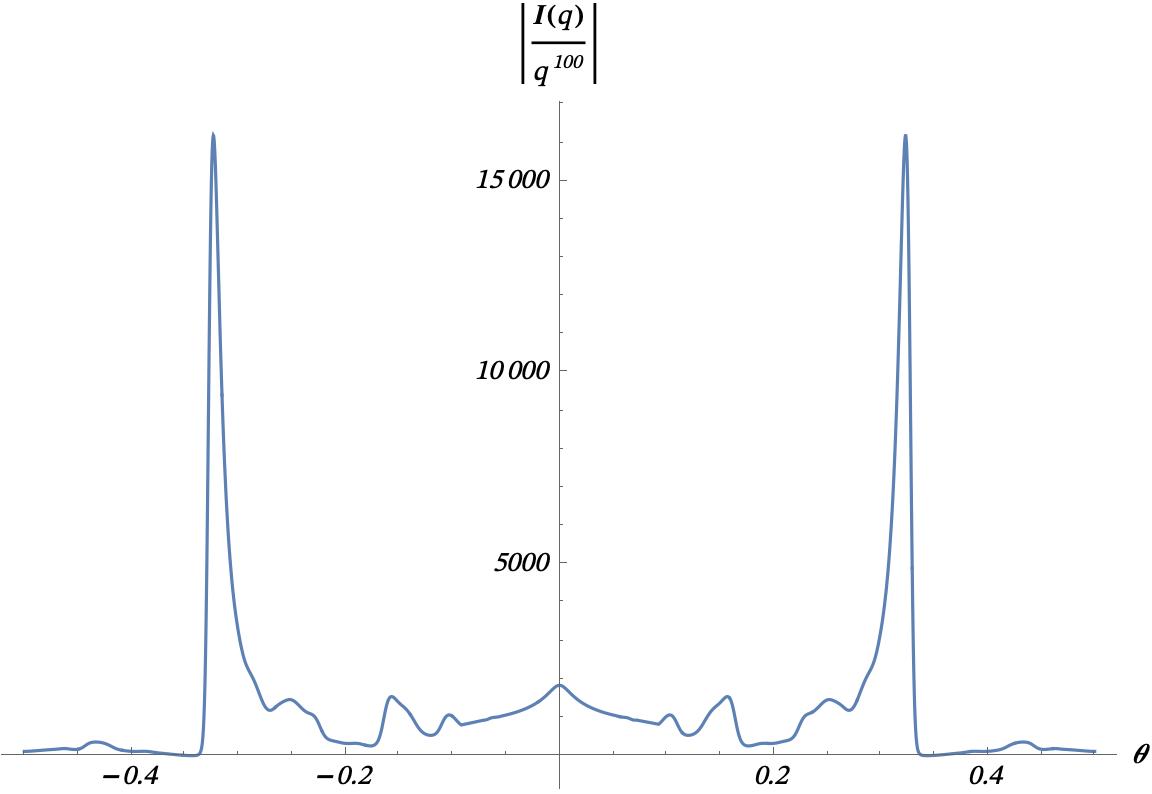}
    \caption{$|\mathcal{I}_{\Phi^{3}}(q)q^{-100}|$ on the contour $q=e^{2 \pi i (\theta +0.01 i)}$ with $\theta\in (-\frac{1}{2},\frac{1}{2})$}.
    \label{100thcoeffcalcphicubedindex}
\end{figure}

Supersymmetric partition functions are expected to be holomorphic functions of $q$ inside the unit circle. Using this as an assumption, we can express the $n$-th coefficient of the $q$-series using the Cauchy integral formula. That is for $Z(q) = \sum_{n=0}^{\infty} a_{n}q^{n}$,
\be \label{cauchyformula1}
a_{n} = \oint \frac{\dd q}{2 \pi i q} q^{-n} Z(q) ,
\ee 
where the contour encloses $q=0$. Near the unit circle, $Z(q)$ diverges or goes to zero. As we take the contour close to the unit circle, the integral \eqref{cauchyformula1} is dominated by saddle points. For example, the integrand that appears in the computation of the $100^{\text{th}}$ coefficient in the $W=\Phi^{3}$ theory has two dominant saddles near $q= e^{\pm \frac{2 \pi i}{3}}$ (see Figure \ref{100thcoeffcalcphicubedindex}).

Supersymmetric partition functions may have singularities \eqref{singWW} at the roots of unity. This aspect is well studied in the context of 3d-3d correspondence where it plays an important role (see {\it e.g.} \cite{Kucharski:2019fgh,Chung:2019jgw}), and is expected to hold more generally. In fact, for a given 3d $\mathcal{N}=2$ theory, different supersymmetric partition functions often exhibit ``universality,'' {\it i.e.} have the same asymptotic behavior \cite{Gadde:2013wq}. Therefore, we expect the saddle points of the integral to be located near the roots of unity. Suppose we have a dominant saddle near $q= e^{-2 \pi i r}$. Using the asymptotic behavior of supersymmetric partition functions from Conjecture \ref{bpscardy2}, we have
\be 
a_{n} \sim \int_{-r- \frac{1}{2}+i \epsilon}^{-r+\frac{1}{2}+i \epsilon} \dd \tau \hspace{0.1cm} \exp[-\frac{1}{\hbar_{r}} \left( \widetilde{\mathcal{W}}_{r}^{(0)}+\hbar_{r} \left(\widetilde{\mathcal{W}}_{r}^{(1)}-2\pi i r n\right)+ \hbar_{r}^{2} \left( \widetilde{\mathcal{W}}_{r}^{(2)} - n \right) \right)]
\ee 
Now the integral has saddle points at $\tau = -r \pm \frac{1}{2 \pi} \sqrt{\frac{\widetilde{W}_{r}^{(0)}}{n-\widetilde{W}_{r}^{(2)}}}$, and the value of integrand at these saddle point is 
$$ \exp(2 \pi i r n - \widetilde{W}_{r}^{(1)} \mp 2 i \sqrt{\widetilde{W}_{r}^{(0)} n - \widetilde{W}_{r}^{(0)}\widetilde{W}_{r}^{(2)}  }).  $$
As $a_{n}$ are real numbers, we expect equally dominant saddle points near $\tau =r$, which have the same real part but opposite imaginary part. Considering only the most dominant saddles, at large $n$ we get,
\be 
a_{n} \sim \mathrm{Re}\left[\exp(\sqrt{\frac{2 \pi^{2}}{3} c_{\text{eff}} \hspace{0.15cm} n }  + 2 \pi i r n )\right],
\ee 
where $c_{\text{eff}} = -\frac{6}{\pi ^{2}} \widetilde{W}_{r}^{(0)}$.

\subsection{Example: \texorpdfstring{$W= \Phi^{3}$}{Phi-cube} theory}

The supersymmetric index of the $W= \Phi^{3}$ theory is given by $\mathcal{I}_{\Phi^{3}}(q) = \frac{(q^{2/3},q)_{\infty}}{(q^{1/3},q)_{\infty}}$. An important detail is that this index has a cubic branch cut from $0$ to $-1$. To avoid working with branch cuts, we change the variables $q \rightarrow q^{3}$. Now the index can be written as
\be 
\mathcal{I}_{\Phi^{3}}(q) = \frac{(q^{2},q^{3})_{\infty}}{(q,q^{3})_{\infty}}. 
\ee 
The index is well-defined inside the unit circle. We can write it as
\begin{equation} \label{phicubedindasympoly}
\mathcal{I}_{\Phi^{3}}(q) =\exp ( \sum_{n=0}^{\infty} \frac{B_{n}}{n!} \left(\log q^{3}\right)^{n-1} (\mathrm{Li}_{2- n}(q^{2})-\mathrm{Li}_{2-n}(q))   ) .
\end{equation}
From the equation \eqref{phicubedindasympoly}, we can see that the index $\mathcal{I}_{\Phi^{3}}(q)$ has singularities at the cubic roots of unity. As $q$ approaches these points, the asymptotic forms of $\mathcal{I}_{\Phi^{3}}(q)$ looks like
\begin{align}
    \mathcal{I}_{\Phi^{3}}(q) &  \sim _{q  \rightarrow 1} \exp( -\frac{1}{3} \log (\hbar_{0} )+O\left(1\right)),\\
    \mathcal{I}_{\Phi^{3}}(q) &  \sim _{q \rightarrow e^{-\frac{2 \pi i}{3}}} \exp( - \frac{\mathrm{Li}_{2}(e^{-\frac{2 \pi i}{3}}) - \mathrm{Li}_{2}(e^{- \frac{4 \pi i }{3}}) }{3 h_{\frac{1}{3}}} + O(1)),\\
    \mathcal{I}_{\Phi^{3}}(q) &  \sim _{q \rightarrow e^{\frac{2 \pi i}{3}}} \exp( - \frac{\mathrm{Li}_{2}(e^{\frac{2 \pi i}{3}}) - \mathrm{Li}_{2}(e^{\frac{4 \pi i }{3}}) }{3 h_{-\frac{1}{3}}} + O(1)).
\end{align}
See appendix \ref{app:asym-qpoch} for details. As expected from Figure \ref{100thcoeffcalcphicubedindex}, the singularities at $q = e^{\pm \frac{2 \pi i }{3}}$ are more dominant compared to that at $q=1$. Therefore, we get dominant contribution from saddle points near $q = e^{\pm \frac{2 \pi i }{3}} $. The effective super-potential $\widetilde{W}_{ \pm \frac{1}{3}}^{(0)} $ is,
\be 
\widetilde{W}_{ \pm \frac{1}{3}}^{(0)}  = \pm \frac{\mathrm{Li}_{2}(e^{\frac{4 \pi i}{3}})-\mathrm{Li}_{2}(e^{\frac{2 \pi i}{3}})}{3}.
\ee 
Hence, for the $W = \Phi^{3}$ theory, the asymptotics of coefficients $a_{n}$ and the effective central charge $c_{\text{eff}}$ are given by 
\begin{align}
        a_{n}  & \sim \mathrm{Re} \left[\exp(\sqrt{\frac{4}{3} \left(\mathrm{Li}_{2}(e^{\frac{2 \pi i}{3}})- \mathrm{Li}_{2}(e^{\frac{4 \pi i}{3}}) \right) \hspace{0.15cm} n }  + \frac{2 \pi i  n}{3} )\right] , \\
        c_{\text{eff}} &= \frac{2}{\pi^{2}}  \left(\mathrm{Li}_{2}(e^{\frac{2 \pi i}{3}})- \mathrm{Li}_{2}(e^{\frac{4 \pi i}{3}}) \right) \approx 0.274227 i . 
\end{align}


\section{Strongly coupled examples: theories \texorpdfstring{$T[M_{3}]$}{T[M3]}}
\label{sec:3d3d}

For intrinsically strongly coupled 3d $\mathcal{N}=2$ theories, we turn to 3d-3d correspondence. This correspondence associates 3d $\mathcal{N}=2$ theories to 3-manifolds via compactification of 6d $(0,2)$ fivebrane theory on 3-manifolds. For a given choice of a 3-manifold, the resulting 3d $\mathcal{N}=2$ theory is usually denoted $T[M_3]$.\footnote{Note, that in early studies of this correspondence, a Lagrangian description for a particular sector of the full theory $T[M_3]$ was proposed \cite{Dimofte2011GaugeTL}, but it will not suffice for our analysis below since abelian flat connections on $M_3$ are not accounted by this sector of $T[M_3]$. See \cite{Chun:2019mal,Chung:2019khu} for a relatively recent account of some of these issues and a more precise characterization of the sector of $T[M_3]$ described by the ``DGG theory.''} In this context, a large supply of 3d $\mathcal{N}=2$ strongly coupled theories comes from the rich world of 3-manifolds. Suppressed in the notation $T[M_3]$ is also a choice of the root system, {\it i.e.} a choice of 6d $(0,2)$ theory. In this paper, we focus mostly on the simplest non-trivial case of $G=SU(2)$ (or, equivalently, $G_{\C} = SL(2,\C)$) that corresponds to two fivebranes in M-theory realization of 3d-3d correspondence. It would be very interesting to extend the analysis below to higher rank version of 3d-3d correspondence and to root systems of Cartan type $D$ and $E$.

Moreover, for 3d $\mathcal{N}=2$ theories that come from 3d-3d correspondence, we have two useful properties that we can leverage in testing Conjectures \ref{bpscardy1} and \ref{bpscardy2}:

\begin{itemize}

\item First, we can relate the superconformal index $\mathcal{I}(q)$ to another BPS partition function that was actively studied in recent years, the so-called half-index or partition function on $D^{2}\times_{q}S^{1}$, where $D^2$ can be understood as the cigar geometry \cite{Gadde:2013wq}:
\be 
\hat{Z}_{a}(q)= Z(D^{2}\times_{q}S^{1}, \mathcal{B}_{a}).
\label{2d3d}
\ee
This partition function depends on a choice of 2d $\mathcal{N}=(0,2)$ boundary condition $\mathcal{B}_{a}$, and can be viewed as the elliptic genus of 2d-3d coupled system. Indeed, if 3d $\mathcal{N}=2$ theory is gapped (or, gappable), then it is just the usual elliptic genus of the 2d $\mathcal{N}=(0,2)$ boundary theory $\mathcal{B}_{a}$. The physics of such coupled 2d-3d systems is very rich and has been an area of active research in recent years. In particular, as we already mentioned above, in the context of 3d-3d correspondence, theories $T[M_3]$ come equipped with a ``canonical'' set of boundary conditions $\mathcal{B}_{a}$ labeled by $a \in \text{Spin}^c (M_3)$. The significance of this insignificant-sounding fact is that $\mathcal{I}(q)$ can be expressed as a sum of more basic building blocks \cite{Gukov:2017kmk}:
\be 
\mathcal{I}_{T[M_{3}]} (q) = \sum_{a} |\mathcal{W}_{a}| \, Z_{T[M_{3}]}(D^{2}\times_{q}S^{1}, \mathcal{B}_{a}) \, Z_{T[\overline{M_{3}}]}(D^{2}\times_{q}S^{1}, \mathcal{B}_{a}),
\label{Idecomp}
\ee 
where the line over $M_{3}$ indicates that $\overline{M_{3}}$ and $M_{3}$ have opposite orientations, and
\be 
Z_{T[M_{3}]}(D^{2}\times_{q}S^{1}, \mathcal{B}_{a}) = \frac{\hat{Z}_{a}(M_{3},q)}{\eta(q)}
\label{ZZZ}
\ee
is the normalized version of \eqref{2d3d}. (For higher-rank version of 3d-3d correspondence, this normalization would involve a factor $\eta (q)^{\text{rank} (G)}$.)

\item The second feature of 3d $\mathcal{N}=2$ theories $T[M_3]$ is that, by the very nature of 3d-3d correspondence, physical properties and observables in $T[M_3]$ are topological invariants of $M_3$. In particular, using the relation \eqref{ZZZ} and the topological interpretation of $\hat{Z}_{a}(q)$ as a $q$-series invariant that provides the non-perturbative definition for complex Chern-Simons theory on $M_3$, we will be able to propose an explicit expression for $c_{\text{eff}}$ in terms of topological invariants of $M_3$. The key element is the trans-series expansion of $\hat{Z}_a (q)$:
\begin{align}
\hat{Z}(q)  &= \sqrt{\frac{\pi i}{\hbar}} \sum_{\mathbb{\bbalpha}\in \pi_{0}(\mathcal{M}_\text{flat}[M_{3},SL(2,\C)])\times \Z} n_{\mathbb{\bbalpha}, 0} \, e^{ \frac{4\pi^2}{\hbar} S_{\mathbb{\bbalpha}}} Z_{\alpha}^{\text{Pert}}(\hbar),  \nonumber \\
&= \sqrt{\frac{\pi i}{\hbar}} \sum_{\alpha \in \pi_{0}(\mathcal{M}_\text{flat}[M_{3},SL(2,\C)])} \sum_{m \in \Z } n_{(\alpha,m), 0} \hspace{0.1cm} \tilde q^{-\mathrm{CS}(\alpha) + m} Z_{\alpha}^{\text{Pert}}(\hbar), \label{trans-series}
\end{align}
writen here for homology spheres and for $G_{\mathbb{C}} = SL(2,\C)$. The derivation of this expansion and its generalization to other 3-manifolds can be found {\it e.g.} in \cite{Gukov:2017kmk,Gukov:2016njj}. The notations used in \eqref{trans-series} and their application to the computation of $c_{\text{eff}}$ will be explained shortly.

\end{itemize}

Since 3d-3d correspondence tells us \cite{Dimofte:2010tz} that the values of twisted superpotential of $T[M_{3}]$ at its critical points are the $SL(2,\C)$ Chern-Simons values of $M_{3}$, one might expect {\it a priori} that the value of $c_{\text{eff}}$ for $T[M_{3}]$ is related to the Chern-Simons values of $M_{3}$. However, for a given $M_3$, there are, in general, many complex flat connections, but there is only one value of $c_{\text{eff}}$. Therefore, a natural question is: If this reasoning is on the right track, {\it which} particular Chern-Simons value of $M_{3}$ determines the value of $c_{\text{eff}}$ for a given $T[M_{3}]$? Or, is it a certain combination of Chern-Simons values? If so, which combination? And, what is the topological significance of complex flat connections that contribute to $c_{\text{eff}}$?

All these questions can be answered with the help of the resurgent analysis, which among other things, leads to \eqref{trans-series}. So, let us discuss this aspect in more detail, starting with the notations. As in \cite{Gukov:2016njj}, we labeled (connected components of the space of) complex flat connections on $M_3$ by
\be
\alpha \in  \pi_{0}\left( \mathcal{M}_{\text{flat}}\left[ M_{3},SL(2,\C) \right]  \right)
\ee
and introduce a notation 
\be
\mathbb{\bbalpha} \in \pi_{0}\left( \mathcal{M}_{\text{flat}}\left[ M_{3},SL(2,\C) \right]  \right) \times \Z
\ee
for their integral lift. In other words, $\alpha$ is an equivalence class of $\mathbb{\bbalpha}$. We follow a general rule that letters from the Greek alphabet denote generic flat connections, whereas letters from the Latin alphabet refer to abelian flat connections. The Chern-Simons value of any $SL(2,\C)$ flat connection is only defined modulo $1$, that is  $\mathrm{CS}(\alpha) \in \C/\Z$. However, for the lift $\mathbb{\bbalpha}$ of $\alpha$, we can define the Chern-Simons value, $S_{\mathbb{\bbalpha}}$, valued in $\C$, such that $S_{\mathbb{\bbalpha}} = \mathrm{CS}(\alpha) \mod 1$.

The coupling constant in (non-perturbative) complex Charn-Simons theory is a continuous complex parameter $q$, the same variable that appears in the superconformal index \eqref{SSindex} of the 3d $\mathcal{N}=2$ theory $T [M_3]$. It is related to the perturbative coupling constant $\hbar$ in the formal power series $Z_{\alpha}^{\text{Pert}}(\hbar)$ via $q = e^{- \hbar}$. In comparison to analytic continuation from Chern-Simons theory with {\it compact} gauge group, it is also sometimes useful to keep in mind the relation to ``level'' $k$, namely $q = e^{-\hbar} = e^{\frac{2\pi i}{k}}$. And another standard notation from complex Chern-Simons theory that we will need is $\tilde q = e^{- 4\pi^2 / \hbar}$. For convenience, we summarize these variables here
\be
q = e^{- \hbar}
\qquad , \qquad
\tilde q = e^{- 4\pi^2 / \hbar}
\ee

The last and, arguably, the most delicate ingredient in the trans-series expansion \eqref{trans-series} is the set of trans-series coefficients $n_{\mathbb{\bbalpha}, 0}$. At present, no general systematic way for computing $n_{\mathbb{\bbalpha}, 0}$ is known, unlike Chern-Simons values that can be computed {\it e.g.} from surgery presentations of $M_3$ and in multiple other ways. Therefore, in our general analysis below, we will not make any assumptions about the values of these coefficients, and when it comes to explicit calculations in concrete examples, we determine their values either from modular properties or numerically.

Now we are ready to combine \eqref{Idecomp}, \eqref{ZZZ}, and \eqref{trans-series} to explore the implications. Suppose that $M_3$ is a homology sphere, and suppose that the most dominant contribution to \eqref{trans-series} near $q=1$ (that is, $\hbar \approx 0$) comes from the term $e^{\frac{4 \pi^2}{\hbar} S_{\mathbb{\bbalpha}}} $. Then, near $q = e^{-\hbar} \approx 1$, the asymptotics of $Z_{T[M_{3}]}(D^{2}\times_{q}S^{1}, \mathcal{B}_{0})$ is given by 
\be 
Z_{T[M_{3}]}(D^{2}\times_{q}S^{1}, \mathcal{B}_{0}) = \frac{\hat{Z}(q)}{\eta(q)}   \sim  \exp[\frac{\pi^{2}}{6 \hbar}\left(1+ 24 S_{\mathbb{\bbbeta}} \right)].
\ee 
Therefore, the growth of the coefficients in $Z_{T[M_{3}]}(D^{2}\times_{q}S^{1}, \mathcal{B}_{0})$, and $c_{\text{eff},\frac{1}{2}}^{T[M_{3}]}$ for the half index are given by
\begin{align}
    a_{n} \sim &  \exp[\sqrt{\frac{2 \pi^{2}}{3} \left[1+24 S_{\mathbb{\bbbeta}} \right] n }], & c_{\text{eff},\frac{1}{2}}^{T[M_{3}]} &= 1 + 24 S_{\mathbb{\bbbeta}}.
\end{align}

Just as we expressed the asymptotic behavior of $Z_{T[M_{3}]}(D^{2}\times_{q}S^{1}, \mathcal{B}_{0})$, we can write the asymptotic behavior of $Z_{T[\overline{M_{3}}]}(D^{2}\times_{q}S^{1}, \mathcal{B}_{0})$ for $\overline{M_{3}}$, ($M_{3}$ with orientation reversal),
\be 
Z_{T[\overline{M_{3}}]}(D^{2}\times_{q}S^{1}, \mathcal{B}_{0})  \sim  \exp[\frac{\pi^{2}}{6 \hbar}\left(1+ 24 S_{\mathbb{\bbbeta}^{\prime}} \right)].
\ee 
Then, the asymptotic behavior of the index is given by 
\be 
\mathcal{I}_{T[M_{3}]} \sim  \exp[\frac{\pi^{2}}{6 \hbar}\left(2+ 24 S_{\mathbb{\bbbeta}^{\prime}}+24 S_{\mathbb{\bbbeta}} \right)].
\ee
and $c_{\text{eff}}$ for $T[M_{3}]$ comes out to be 
\be
c_{\text{eff}} \; = \; 2+24 S_{\mathbb{\bbbeta}}+ 24 S_{\mathbb{\bbbeta}^{\prime}}.
\label{ceff-trna-series}
\ee
Given that $M_{3}$ and $\overline{M_{3}}$ differ by the orientation reversal, one might expect $S_{\mathbb{\bbbeta}}$ and $S_{\mathbb{\bbbeta}^{\prime}}$ to cancel each other, resulting in $c_{\text{eff}}=2$. However, as we will see shortly, this does not happen, even in simple examples. Moreover, it is important to stress that $c_{\text{eff}}$ is determined not only by Chern-Simons values, but also by trans-series coefficients or, equivalently, by values of $\mathbb{\bbbeta}$ and $\mathbb{\bbbeta}^{\prime}$ that appear in \eqref{ceff-trna-series}. It is this latter data that is hard to determine {\it a priori}, and which certainly deserves to be studied more systematically.


\subsection{Examples: \texorpdfstring{$M_3 = \Sigma(2,3 , 6 \pm 1)$}{(2,3,5) and (2,3,7) Brieskorn spheres}}

Let us consider an example of $T[M_{3}]$ for $M_{3} = \Sigma(2,3,5)$. The three-manifold $\Sigma(2,3,5)$ has three $SL(2,\C)$ flat connections, one abelian connection, and two non-abelian connections. We denote the abelian connection by $\alpha_{0}$ and the two non-abelian connections by $\alpha_{1}$ and $\alpha_{2}$. The Chern-Simons value of the abelian connection is zero, while the Chern-Simons values of non-abelian connections are given by 
\begin{align}
    \mathrm{CS}(\alpha_{1})&= -\frac{1}{120} \mod 1  & \mathrm{CS}(\alpha_{1})&= -\frac{49}{120} \mod 1.
\end{align}
For the three-manifold $\overline{\Sigma(2,3,5)}$ with opposite orientation, we have the corresponding counterparts of these three connections, which we denote by $\alpha_{0}^{\prime}, \alpha_{1}^{\prime}$, and $\alpha_{2}^{\prime}$. Since $\overline{\Sigma(2,3,5)}$ has opposite orientation compared to that of $\Sigma(2,3,5)$, the Chern-Simons values of $\alpha_{0}^{\prime}, \alpha_{1}^{\prime}$, and $\alpha_{2}^{\prime}$ are also opposite:
\begin{align}
   \mathrm{CS}(\alpha_{0}^{\prime})&= 0 \mod 1  & \mathrm{CS}(\alpha_{1}^{\prime})&= \frac{1}{120} \mod 1  & \mathrm{CS}(\alpha_{1})&= \frac{49}{120} \mod 1.
\end{align}

The $\hat{Z}(q)$ invariant for $M_{3}= \overline{\Sigma(2,3,5)}$ is related to the order-five mock theta function $\chi_{0}(q)$ in the following way,
\be 
 \hat{Z}(\overline{\Sigma(2,3,5)};q) = q^{\frac{3}{2}} \chi_{0}(q) = q^{\frac{3}{2}} \sum_{n=0}^{\infty} \frac{q^{n}}{(q^{n+1})_{n}}.
\ee 
The modular properties of this mock theta function have been studied in \cite{Zwegers2008MockTF, Gordon2003}. The modular transform of $\chi_{0}(q)$ is given by, 
\begin{align}
    q^{-\frac{1}{120}}(\chi_{0}(q)-2) =&  - \sqrt{ \frac{\pi}{\hbar}}  \sqrt{\frac{5-\sqrt{5}}{5}} \tilde q^{-\frac{1}{120}} (\chi_{0}(\tilde q)-2) - \sqrt{ \frac{\pi}{\hbar}}  \sqrt{\frac{5+\sqrt{5}}{5}} \tilde q^{-\frac{49}{120}+1} \chi_{1}(\tilde q) \nonumber \\ & \hspace{1cm}  -  \sqrt{\frac{135 \hbar}{2 \pi} } J\left(\frac{1}{5}, 5\hbar \right),
\end{align}
where $\tilde q = e^{- 4 \pi^2 / \hbar}$, $\chi_{1} (q)$ is another order-five mock theta function given by 
\be 
\chi_{1} (q)= \sum_{n=0}^{\infty} \frac{q^{n}}{(q^{n+1})_{n+1}}, 
\ee 
and the function $J (r, \gamma )$ is defined as 
\be 
J(r, \gamma  ) = \int_{0}^{\infty} e^{-\frac{3}{2}\gamma  x^{2}} \frac{\cosh[(3r-2)\gamma  x ]+\cosh[(3r-1)\gamma  x ]}{\cosh[\frac{3}{2}\gamma  x ]} \dd x.
\ee 
Using the mock-modularity of $\chi_{0}(q)$, we can write down the mock-modular properties of $\hat{Z}(\overline{\Sigma(2,3,5)};q)$ as follows,
\be \label{modularityof235}
\hat{Z}(\overline{\Sigma(2,3,5)};q)  =  \sqrt{ \frac{\pi i}{\hbar}}  \sum_{n=0}^{\infty} \tilde{a}_{1}(n,q) \tilde q^{-\frac{1}{120}+n} +  \sqrt{ \frac{\pi i}{\hbar}}  \sum_{n=1}^{\infty} \tilde{a}_{2}(n,q) \tilde q^{-\frac{49}{120}+n}  + Z_{0}^{\text{Pert}}(\hbar),
\ee 
where $Z_{0}^{\text{Pert}}(\hbar)$ is,
\be 
Z_{0}^{\text{Pert}}(\hbar) = 2q^{\frac{3}{2}} - \sqrt{\frac{135 \hbar}{2 \pi} } q^{\frac{181}{120}} J\left(\frac{1}{5}, 5 \hbar \right),
\ee 
and the coefficients $\tilde{a}_{1}(n,q)$, and $\tilde{a}_{2}(n,q)$ are,
\begin{align}
    \tilde{a}_{1}(n,q)&= i q^{\frac{181}{120}}   \sqrt{i\frac{5-\sqrt{5}}{5}} a_{1}(n) &
     \tilde{a}_{2}(n,q)&=  i q^{\frac{181}{120}}   \sqrt{i\frac{5+\sqrt{5}}{5}} a_{2}(n)
\end{align}
where $a_{1}(n)$ and $a_{2}(n)$ are $n$-th coefficients of the $q$-series $(\chi_{0}(q)-2)$  and $\chi_{1}(q)$, respectively. Notice that the first sum in equation \eqref{modularityof235} starts from $n=0$, while the second sum starts from $n=1$. From this, we can deduce that the most dominant singularity of $\hat{Z}(\overline{\Sigma(2,3,5)};q) $ on the unit circle is at $q=1$ and is due to the term $\tilde{a}_{1}(0,q) \tilde{q}^{-\frac{1}{120}}$. The asymptotic behavior of the half-index $Z_{T[\overline{\Sigma(2,3,5)}]}(D^{2}\times_{q}S^{1}, \mathcal{B}_{0})$ near $q=1$ is, 
\be 
Z_{T[\overline{\Sigma(2,3,5)}]}(D^{2}\times_{q}S^{1}, \mathcal{B}_{0}) \sim \tilde{q}^{-\frac{1}{24} - \frac{1}{120}}.
\ee 
Therefore, $c_{\text{eff}}$ for the half-index $Z_{T[M_{3}]}(D^{2}\times_{q}S^{1}, \mathcal{B}_{0})$, with $M_{3} =\overline{\Sigma(2,3,5)}$, is
\be 
c_{\text{eff},\frac{1}{2}}^{T[\overline{\Sigma(2,3,5)}]} = \frac{6}{5}.
\ee 

The $\hat{Z}(q)$ invariant for $M_{3}= \Sigma(2,3,5)$ is given by 
\be 
\hat{Z}(\Sigma(2,3,5),q) = q^{-1}\sum_{n=0}^{\infty} \frac{(-1)^{n}q^{\frac{n(3n-1)}{2}}}{(q^{n+1})_{n}} = q^{-1}\left( 2- q^{-\frac{1}{120}} \Psi_{30}^{(1)+(11)+(19)+(29)} (q)\right),
\ee 
where $\Psi_{30}^{(1)+(11)+(19)+(29)} (q)$ is a false theta function. Here we have used a shorthand notation 
\be 
\Psi_{p}^{(a_{1}) +(a_{2}) +\cdots }(q) = \Psi_{p}^{(a_{1})}(q)+\Psi_{p}^{(a_{2})}(q)+\cdots.
\ee 
The false theta functions $\Psi_{p}^{(a)}$ are given by 
\be 
\Psi_{p}^{(a)}(q) = \sum_{n \in 2 p \Z +a } \mathrm{sign}(n)q^{\frac{n^{2}}{4 p }}. 
\ee 
The modular transform of the false theta function $\Psi_{p}^{(a)}(q)$ is given by
\be 
\Psi_{p}^{(a)} (q) = - \sqrt{- \frac{2 \pi}{\hbar} } \sum_{b=1}^{p-1} M_{ab} \Psi_{p}^{(b)} (\tilde q) + \sum_{n=0}^{\infty} c_n \hbar^n \left( \frac{-1}{4 p} \right)^{n},
\ee 
where $\sum_{n=0}^{\infty} c_{n} \hbar^n \left(\frac{-1}{4 p}\right)^{n}$ is given by the perturbative expansion of partition function of Chern-Simons around the abelian flat connection, and the matrix $M_{ab}$ is given by 
\be 
M_{ab} = \sqrt{\frac{2}{p}} \sin{\frac{\pi a b}{p}}.
\ee 
Using the false-modularity of false theta functions, we get the modular transform of the $\hat{Z}(\Sigma(2,3,5),q)$,
\be 
\hat{Z}(\Sigma(2,3,5),q) =  \sqrt{\frac{\pi i}{\hbar}}  \sum_{n=0}^{\infty} \tilde{b}_{1}(n,q) \tilde q^{\frac{1}{120}+n} + \sqrt{\frac{\pi i}{\hbar}}  \sum_{n=0}^{\infty} \tilde{b}_{2}(n,q) \tilde q^{\frac{49}{120}+n} + Z_{0}^{\text{Pert}} (\hbar),
\ee 
where the coefficients $\tilde{b}_{1}(n,q)$ and $\tilde{b}_{2}(n,q)$ are, 
\begin{align}
    \tilde{b}_{1}(n,q) &= -i q^{-\frac{121}{120}}\sqrt{\frac{5-\sqrt{5}}{5i}} b_{1}(n), &  \tilde{b}_{1}(n,q) &= -i q^{-\frac{121}{120}}\sqrt{\frac{5+\sqrt{5}}{5i}} b_{2}(n),
\end{align}
where $b_{1}(n)$, and $b_{2}(n)$ are the coefficients of the $q$-series $q^{-\frac{1}{120}} \Psi_{30}^{(1)+(11)+(19)+(29)}(q)$, and  $q^{-\frac{49}{120}} \Psi_{30}^{(7)+(13)+(17)+(23)}(q)$, respectively. The first few terms of these $q$-series are 
\begin{align}
     q^{-\frac{1}{120}}\Psi_{30}^{(1)+(11)+(19)+(29)}(q) &= 1+  q +  q^{3} + q^{7} -q^{8}+\cdots , \\
     q^{-\frac{49}{120}} \Psi_{30}^{(7)+(13)+(17)+(23)} ( q) &= 1 +  q +  q^{2} +  q^{4} - q^{11} +\cdots.
\end{align}
Therefore, the dominant term governing the asymptotic behavior of $\hat{Z}(\Sigma(2,3,5),q)$ near $q=1$ is $Z_{0}^{\text{Pert}} (\hbar)$. Thus, the asymptotic behavior of the half-index $Z_{T[\Sigma(2,3,5)]}(D^{2}\times_{q}S^{1}, \mathcal{B}_{0})$ near $q=1$ is, 
\be 
Z_{T[\Sigma(2,3,5)]}(D^{2}\times_{q} S^{1}, \mathcal{B}_{0}) \sim \tilde q^{-\frac{1}{24}},
\ee 
and the asymptotic behavior of the index $\mathcal{I}_{T[\Sigma(2,3,5)]}$ near $q=1$ is,
\be 
\mathcal{I}_{T[\Sigma(2,3,5)]}(q) \sim \tilde q^{-\frac{11}{120}}.
\ee
Therefore the $c_{\text{eff}}$ of the index of $T[\Sigma(2,3,5)]$ is 
\be 
c_{\text{eff}}^{T[\Sigma(2,3,5)]} = \frac{11}{5}.
\ee 

The $\hat{Z}(q)$ invariant for $\overline{\Sigma(2,3,7)}$ is related to the order-seven mock theta function $\mathcal{F}_{0}(q)$ as follows,
\be 
\hat{Z}(\overline{\Sigma(2,3,7)};q) = q^{-\frac{1}{2}} \mathcal{F}_{0}(q) = q^{-\frac{1}{2}} \sum_{n=0}^{\infty} \frac{q^{n^{2}}}{(q^{n+1})_{n}}.
\ee 
Just as we used the mock-modularity of order-five mock theta function $\chi_{0}(q)$ to determine $c_{\text{eff}}$ for half-index of $T[\overline{\Sigma(2,3,5)}]$. We can use the mock-modularity of $\mathcal{F}_{0}(q)$ to get the asymptotic behavior of the half-index of $T[\overline{\Sigma(2,3,7)}]$ near $q=1$ and the $c_{\text{eff}}$ for half-index of $T[\overline{\Sigma(2,3,7)}]$. They are given by \begin{align}
Z_{T[\overline{\Sigma(2,3,7)}]}(D^{2}\times_{q}S^{1}, \mathcal{B}_{0}) &\sim \tilde q^{-\frac{1}{21}},     &     c_{\text{eff},\frac{1}{2}}^{T[\overline{\Sigma(2,3,7)}]} &= \frac{8}{7}.
\end{align}

Just as for $\Sigma(2,3,5)$, the $\hat{Z}$-invariant of $\Sigma(2,3,7)$ is a linear combination of the false theta functions $\Psi_{168}^{(a)}$. The asymptotic behavior of $\hat{Z}(\Sigma(2,3,7),q)$ near $q=1$ is governed by  $Z_{0}^{\text{Pert}}(\hbar)$. Thus the asymptotic behavior of the index of $T[\Sigma(2,3,7)]$ near $q=1$ and the $c_{\text{eff}}$ of the index of $T[\Sigma(2,3,7)]$ are given by 
\begin{align}
 \mathcal{I}_{T[\Sigma(2,3,7)]}(q) &\sim  \tilde q^{-\frac{5}{56}}, &  c_{\text{eff}}^{T[\Sigma(2,3,7)]}  &= \frac{15}{7}.
\end{align}


\subsection{Numerical estimates for \texorpdfstring{$T[\Sigma(s,t,s t - 1)]$}{(s,t,st-1) Brieskorn spheres}}

Small surgeries on torus knots give us a special class of Brieskorn homology spheres:
\begin{align*}
   S^{3}_{\frac{1}{r}} \left( T(s,t)  \right)  & =  \overline{\Sigma(s,t , r s t - 1)} &  S^{3}_{\frac{1}{r}} \left( T(s,-t)  \right)  & =  \overline{\Sigma(s,t , r s t + 1)} \\ 
   S^{3}_{-\frac{1}{r}} \left( T(s,-t)  \right)  & =  \Sigma(s,t , r s t - 1) &  S^{3}_{-\frac{1}{r}} \left( T(s,t)  \right)  & =  \Sigma(s,t , r s t + 1).
\end{align*}
Using the regularised surgery formula from \cite{Park:2021ufu} and the two-variable series, $F_{K}(x,q)$, for torus knots from \cite{Gukov2019ATS}, we can write down closed form formulae for $\hat{Z} (\Sigma(s,t , r s t \pm 1),q)$ and $\hat{Z}(\overline{\Sigma(s,t , r s t \pm 1)},q)$ as follows:
\begin{align}
    \hat{Z}(\Sigma(s,t , r s t \pm  1),q) & = q^{-\frac{r+r^{-1}}{4}} \sum_{j = 0}^{\infty} \varepsilon_{2j+1}q^{ \pm  \frac{j(j+1)}{s t} \pm \frac{(t^{2}-1)(s^{2}-1)}{4 s t}}  \left(q^{r(j+\frac{1}{2}-\frac{1}{2r})^{2}}-q^{r(j+\frac{1}{2}+\frac{1}{2r})^{2}}\right),
\end{align}
\begin{align}\label{prsurgeryontorusknots}
    \hat{Z}(\overline{\Sigma(s,t , r s t \pm  1)},q) & = q^{\frac{r+r^{-1}}{4}} \sum_{j = 0}^{\infty} \varepsilon_{2j+1}q^{ \mp  \frac{j(j+1)}{s t} \mp  \frac{(t^{2}-1)(s^{2}-1)}{4 s t}}  \left(q^{-r(j+\frac{1}{2}-\frac{1}{2r})^{2}}-q^{-r(j+\frac{1}{2}+\frac{1}{2r})^{2}}\right) \endline & \hspace{3cm} \times  \left(1-\frac{\sum_{|k|\leq j} (-1)^{k} q^{\frac{k((2r+1)k+1)}{2}}  }{(q^{r},q^{2r+1})_{\infty} (q^{r+1},q^{2r+1})_{\infty} (q^{2r+1},q^{2r+1})_{\infty} } \right).
\end{align}
The $q$-series, $\hat{Z}(\Sigma(s,t , r s t \pm  1),q)$ has coefficients valued in $\{\pm1,0\}$. It is a linear combination of false theta functions $\Psi_{p}^{(a)}(q)$. Similar to $\Sigma(2,3,5)$, the asymptotics of $\hat{Z}(\Sigma(s,t , r s t \pm  1),q)$ near $q=1$ are dominated by $Z_{0}^{\text{Pert}}$. Therefore, the asymptotic behavior of $Z_{T[\Sigma(s,t , r s t \pm  1)}(D^{2}\times_{q}S^{1}, \mathcal{B}_{0}) $ near $q=1$ is,
\be 
Z_{T[\Sigma(s,t , r s t \pm  1)}(D^{2}\times_{q}S^{1}, \mathcal{B}_{0}) \sim \tilde q^{-\frac{1}{24}}.
\ee 
On the other hand, the coefficients of the $q$-series, $\hat{Z}(\overline{\Sigma(s,t , r s t \pm  1)},q)$ grow rapidly. From the trans-series representation of $\hat {Z}(q)$, we expect the asymptotic behavior of $\hat{Z}(\overline{\Sigma(s,t , r s t \pm  1)},q)$ near $q=1$ to be of the form, $\hat{Z}(\overline{\Sigma(s,t , r s t \pm  1)},q) \sim \tilde q^{-S_{\mathbb{\bbalpha}}}$, for some non-abelian flat connection $\alpha$. The Chern-Simons values of flat connections on $\overline{\Sigma(s,t , r s t \pm  1)}$ are of the form
\be 
\mathrm{CS}(\alpha_{n}) = \frac{m^{2}}{4 s t (r s t \pm 1)} \mod 1, \hspace{2cm} m \in \Z. 
\ee 
Thus, the expected form of trans-series representation of $\hat{Z}(q)$ tells us that asymptotically, near $q=1$,
\be 
\hat{Z}(\overline{\Sigma(s,t , r s t \pm  1)},q) \sim \tilde q^{- \frac{m^{2}}{4 s t (r s t \pm 1)} + \ell },
\ee 
for some integers $m,\ell$. This, in turn, tells us that the asymptotic coefficients of the $q$-series are 
\be 
a_{n}(r,s,t) \sim \exp[\sqrt{b n }] \sim  \exp[\sqrt{16 \pi^{2} \left(\frac{m^{2}}{4 s t (r s t \pm 1)} - \ell  \right)n }]
\ee 

\begin{figure}[ht]
     \centering
     \begin{subfigure}[b]{0.4\textwidth}
         \centering
         \includegraphics[width=\textwidth]{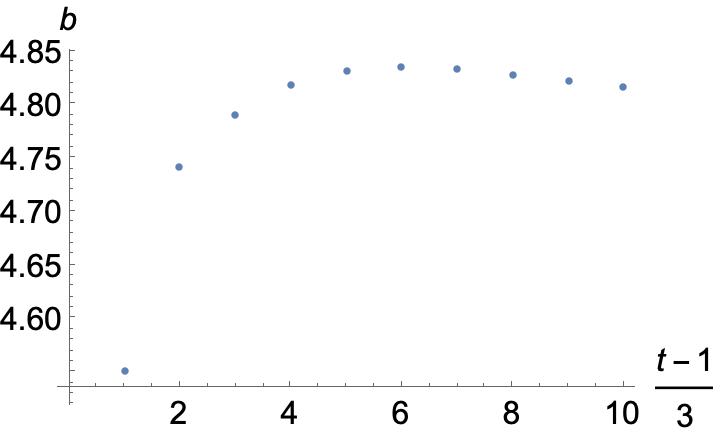}
         \caption{$T(3,t=3x+1)$}
         \label{bsfortorusknots31}
     \end{subfigure}
     \hfill
     \begin{subfigure}[b]{0.4\textwidth}
         \centering
         \includegraphics[width=\textwidth]{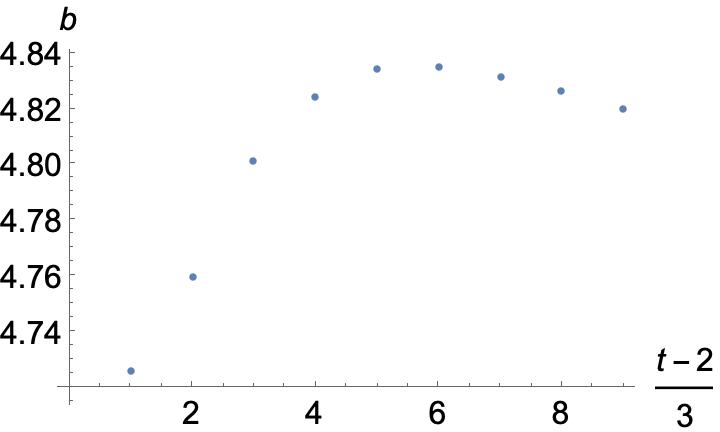}
         \caption{$T(3,t=3x+2)$}
         \label{bsfortorusknots32}
     \end{subfigure}
     \hfill
     \begin{subfigure}[b]{0.5\textwidth}
         \centering
         \includegraphics[width=\textwidth]{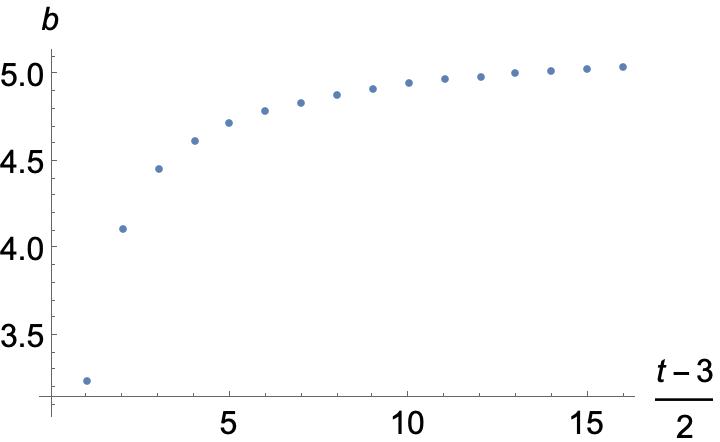}
         \caption{$T(2,t=2x+3)$}
         \label{bsfortorusknots2}
     \end{subfigure}
        \caption{Numerical estimates for $b$ for torus knots of the form $T(2,t=2x+1)$, $T(3,t=3x+1)$, and $T(3,t=3x+2)$.}
        \label{bsfortorusknots}
\end{figure} 

One needs the precise modular properties or, equivalently, the trans-series coefficients in order to determine the integers $m$ and $\ell$. We use the closed form formula \eqref{prsurgeryontorusknots} and study the growth of coefficients $a_{n}(s,t)$ to get numerical estimates for the integers $m$ and $\ell$ for $+1$ surgery on various torus knots. In all the examples of torus knots we looked at, the integer $\ell$ is zero. We have plotted the values of $b$ for torus knots of the form $T(2,t=2x+3)$, $T(3,t=3x+1)$, and $T(3,t=3x+2)$ in Figure \ref{bsfortorusknots}. The numerical estimates for $m$ are listed in the table below.
\begin{center}
    \begin{tabular}{|c||c|c|c|}
    \hline
    $t \downarrow$ $s\rightarrow$ & 2 & 3 &  4 \\
    \hline 
     4& &3.90 &  \\
    \hline
     5& 2.72 & 5.01 & 6.69  \\
    \hline
    6& & &  \\
    \hline
   7 & 4.35 & 7.10  & 9.54  \\
    \hline
   8 & &8.16 &  \\
    \hline
   9 & 5.88 & & 12.38  \\
    \hline
   10 & &10.27 &  \\
    \hline
   11 & 7.36 & 11.33 & 15.21  \\
    \hline
   12 & & &  \\
    \hline
   13 & 8.22 & 13.45 & 18.02 \\
    \hline
\end{tabular}
\end{center}

In other words, in this family of examples, we have
\be 
c_{\text{eff}} = 2 + \frac{24 m^{2}}{4 s t (r s t \pm 1)} 
\ee
Note, in our earlier analysis, we determined the exact value $m=1$ for $\Sigma (2,3,6 \pm 1)$.

\section*{Acknowledgements}
We would like to thank Miranda C. N. Cheng, Boris Feigin, Kathrin Bringmann, John Cardy, Angus Gruen, Antun Milas, Piotr Kucharski, Sunghyuk Park, Du Pei, Silviu Pufu, and Nicolai Reshetikhin for helpful discussions.
This work is supported by a Simons Collaboration Grant on New Structures in Low-Dimensional Topology, by the NSF grant DMS-2245099, and by the U.S. Department of Energy, Office of Science, Office of High Energy Physics, under Award No. DE-SC0011632.

\appendix

\section{\texorpdfstring{$q$}{q}-Pochhammer symbols near the unit circle}\label{app:asym-qpoch}

In this appendix, we write down a formula for the logarithm of the q-Pochhammer symbol $\log[(x,q)_{\infty}]$ in the domain $|x|<1$ and $|q|<1$, and use it to derive asymptotics of $(q^{a},q^{b})$ near roots of unity. 
\begin{align}
    \log[(x,q)_{\infty}] & = \sum_{k=0}^{\infty} \log[1-x q^{k}] = -\sum_{k=0}^{\infty} \sum_{\ell=1}^{\infty} \frac{x^{\ell}}{\ell} q^{\ell k} = -  \sum_{\ell=1}^{\infty} \frac{x^{\ell}}{\ell}  \sum_{k=0}^{\infty} q^{\ell k} , \endline
    &= \sum_{\ell=1}^{\infty} \frac{x^{\ell}}{\ell } \frac{1}{q^{\ell}-1} = \sum_{\ell=1}^{\infty} \frac{x^{\ell}}{\ell} \sum_{n=0}^{\infty} \frac{B_{n}}{n!} \left(\log[q^{\ell}]\right)^{n-1} , \endline &= \sum_{n=0}^{\infty}  \frac{B_{n}}{n!} \left(\log[q]\right)^{n-1} \sum_{\ell=1}^{\infty} x^{\ell} \ell^{n-2}, \endline 
    \log[(x,q)_{\infty}] & = \sum_{n=0}^{\infty}  \frac{B_{n}}{n!} \left(\log[q]\right)^{n-1} \mathrm{Li}_{2-n}(x).
\end{align}
Here $B_{n}$ are the Bernoulli numbers, given by the following generating series
\be 
\frac{t }{e^{t}-1} = \sum_{n=0}^{\infty} B_{n} \frac{t^{n}}{n!} .
\ee 
If $r\in \Z$ is such that $a r \neq 0 \mod b$, then we can write the asymptotics of $(q^{a},q^{b})$ as $q\rightarrow e^{\frac{2 \pi i r}{b}}$,
\be 
(q^{a},q^{b}) \sim \exp(\frac{\mathrm{Li}_{2}(q^{a})}{\log(q^{b})} )
\qquad
\text{as}
\quad
q \rightarrow e^{\frac{2 \pi i r}{b}}. 
\ee

\newpage

\bibliographystyle{unsrt}
\bibliography{Bibliography}

\end{document}